\documentclass[floatfix,aps,prd,twocolumn,showpacs,superscriptaddress,nofootinbib]{revtex4} 

\usepackage{graphicx}

\usepackage{amsmath}
\usepackage{amssymb}

\font\FermiSmallfont=cmssq8 scaled 1200

\def\LANLppthead#1#2#3{
\null 
\begin{center}\vskip -1.0truein{\hbox to 7.5truein {
\hfill
\vbox to 1in {\vfill \FermiSmallfont
              \hbox{#1}
              \hbox{#2}
              \hbox{#3}
              \vfill}
}}\vskip-0.0truein\end{center}}

\def\chandra{\emph{Chandra}}
\def\Hubble{\emph{Hubble}}
\def\cmsq{cm$^2$}

\def\msun{M_\odot}

\def\gcmsq{g$\,$cm$^{-2}$}
\def\deg{$^\circ$}

\begin{document}

\LANLppthead {UMD-PP-06-057}{LA-UR 06-3820}{astro-ph/0611144}

\title{ Limits on the Radiative Decay of Sterile Neutrino Dark Matter from 
the Unresolved Cosmic and Soft X-ray Backgrounds}

\author{Kevork N.\ Abazajian}
\affiliation{ Department of Physics, University of Maryland, College
Park, MD 20742, USA} 
\affiliation{Theoretical
  Division, Los Alamos National Laboratory, Los Alamos, New Mexico
  87545, USA} 
\author{Maxim Markevitch}
\affiliation{Harvard-Smithsonian Center for
  Astrophysics, 60 Garden Street, Cambridge, Massachusetts 02138, USA}
\affiliation{Space Research Institute, Russian Acad. Sci., 
  Profsoyuznaya 84/32, Moscow 117997, Russia}
\author{Savvas M.\ Koushiappas}
\affiliation{Theoretical
  Division, Los Alamos National Laboratory, Los Alamos, New Mexico
  87545, USA} 
\affiliation{ISR
  Division, Los Alamos National Laboratory, Los Alamos, New Mexico
  87545, USA} 
\author{Ryan C.\ Hickox}
\affiliation{Harvard-Smithsonian Center for
  Astrophysics, 60 Garden Street, Cambridge, Massachusetts 02138, USA}


\begin{abstract}
  We present upper limits on line emission in the Cosmic X-ray
  background (CXB) that would be produced by decay of sterile neutrino
  dark matter. We employ the spectra of the unresolved component of
  the CXB in the {\it Chandra} Deep Fields North and South obtained with the
  {\it Chandra} CCD detector in the $E=0.8-9$ keV band. The expected decay
  flux comes from the dark matter on the lines of sight through the
  Milky Way galactic halo. Our constraints on the sterile neutrino
  decay rate are sensitive to the modeling of the Milky Way halo. The
  highest halo mass estimates provide a limit on the sterile neutrino
  mass of $m_s<2.9$ keV in the Dodelson-Widrow production model, while
  the lowest halo mass estimates provide the conservative limit of
  $m_s<5.7$ keV ($2\sigma$).  We also discuss constraints from a short
  observation of the softer ($E<1$ keV) X-ray background with a
  rocket-borne calorimeter by McCammon and collaborators.
\end{abstract}

\maketitle

\section{Introduction}

The abundance of cosmological dark matter is now well quantified by
cosmological observations to better than 10\% in density
\cite{Spergel:2006hy,Tegmark:2006az}, yet its identity remains
unknown.  Several indicators point towards a modification of the
properties of dark matter clustering from the cold dark matter (CDM)
paradigm to resolve potential discrepancies that persist with the
ansatz of an absolutely {\em cold} dark matter candidate.  Among
potential problems with CDM are indications for density cores in local
group dwarf spheroidal
galaxies~\cite{Goerdt:2006rw,Sanchez-Salcedo:2006fa,Strigari:2006ue,Gilmore:2006iy}
and the lack of a correspondence of the observed dwarf galaxies with
the number of halos expected with the same maximal velocity dispersion
or mass~\cite{Klypin:1999uc,Moore:1999wf}.  This has prompted the
investigation of alternatives to CDM, such as warm dark matter
(WDM)~\cite{Blumenthal:1982mv,Bode:2000gq}, late-decaying superWIMP
dark matter~\cite{Kaplinghat:2005sy,Cembranos:2005us}, fuzzy cold dark
matter~\cite{Hu:2000ke}, or meta-CDM~\cite{Strigari:2006jf}.

A leading particle candidate for WDM is a fermion that has no standard
model interactions, yet couples with the standard model neutrino
sector via the neutrino mass generation mechanism, namely a sterile
neutrino.  Extensions to the standard model of
particle physics typically include sterile neutrinos, including
left-right symmetric (mirror) models~\cite{Berezhiani:1995yi},
supersymmetric axinos as sterile neutrinos~\cite{Chun:1999cq},
superstring models~\cite{Langacker:1998ut}, models with large extra
dimensions~\cite{Arkani-Hamed:1998vp,Abazajian:2000hw}, and
phenomenological models such as the $\nu$MSM~\cite{Asaka:2005an}.
Other motivations for a sterile neutrino to have parameters in the
parameter space of interest for their oscillation production as dark
matter are the generation of pulsar
kicks~\cite{Kusenko:1998bk,Fuller:2003gy,Kusenko:2004mm}, Type II
supernova shock heating enhancement~\cite{Hidaka:2006sg}, and enhanced
molecular hydrogen formation at high redshift~\cite{Biermann:2006bu}.

Sterile neutrinos may be produced in the early universe via
non-resonant Dodelson-Widrow (DW) oscillation
production~\cite{Dodelson:1993je}, or via a resonant oscillation
production in cosmologies with a non-zero lepton
number~\cite{Shi:1998km}.  Both non-resonant DW and resonant
production models fall in a parameter space that is continuous in the
cosmological lepton number, with non-negligible lepton numbers
allowing for resonant production at smaller mixing angles.  The DW
model is the simplest model for sterile neutrino dark matter
production, because it assumes a standard thermal history in the early
universe, zero lepton numbers, and no additional couplings of the
sterile neutrino.  For the non-resonant and resonant cases, production
occurs near the quark-hadron transition, and the relation between the
critical density in sterile neutrinos and its mixing parameters with
active neutrinos is given by
Refs.~\cite{AbazajianProduction05,Abazajian:2002yz,Abazajian:2001nj}.
Sterile neutrinos can be produced at smaller mixing angles than the DW
mechanism through the resonant production model or via additional
couplings to other particles, such as the
inflaton~\cite{Shaposhnikov:2006xi}.  They could be produced at larger
mixing angles than the DW model if over-abundance due to the
nonresonant oscillation production is avoided through dilution by
massive particle decay after production~\cite{Asaka:2006ek} or by a
low reheating temperature scale in the early
universe~\cite{Gelmini:2004ah}.

A particularly interesting particle mass range for the sterile
neutrino can be framed by the mass required to produce a constant
density core of 100--300 pc like that discussed for the Fornax dwarf
spheroidal~\cite{Goerdt:2006rw,Sanchez-Salcedo:2006fa}, which is
0.5--1.3 keV~\cite{Strigari:2006ue,Abazajian:2006yn}, or more massive
if the phase-space packing limit is not achieved.  In addition, the
requirement for a viable sterile neutrino dark matter candidate
constrains their parameter space by demanding that: 1) the total decay
time-scale is larger than the age of the universe, 2) it is consistent
with radiation energy density constraints from primordial
nucleosynthesis and the cosmic microwave background, 3) it is
consistent with lithium photoproduction constraints, and 4) it is
consistent with the diffuse X-ray background, and X-ray observations
of of clusters or other nearby
structures~\cite{Abazajian:2001vt,Dolgov:2000ew,Drees:2000wi}.
Sterile neutrinos are constrained by X-ray observations because of the
considerable radiative decay width to a photon and lighter-mass
neutrino due to the same coupling required for their production,
producing a spectral line in the X-ray~\cite{Pal:1981rm}.

Abazajian, Fuller \& Tucker~\cite{Abazajian:2001vt} found that X-ray
observations could present the most stringent constraints in the upper
particle-mass range regions of parameter space.  An estimate in that
work of the sensitivities of X-ray observations of the large dark
matter overdensities present in clusters of galaxies and field
galaxies was made and led to an estimated limit from {\em XMM-Newton}
observations of the Virgo cluster on the non-resonant DW production
model on the particle mass of the sterile neutrino.
Recent work has shown that the local radiative decay flux from dark
matter overdensities in Local Group structures can be comparable to
that estimated from clusters of galaxies and field galaxies, with
reduced continuum emission and bigger angular size, and therefore
increased sensitivity~\cite{Boyarsky:2006fg}.  This was also found in
subsequent observational work
\cite{Watson:2006qb,Boyarsky:2006ag,Riemer-Sorensen:2006fh}.  For
example, using a nondetection of the decay line from the Andromeda
galaxy by {\em XMM-Newton}, Watson et al.~\cite{Watson:2006qb} derived
$m_s<3.5$ keV in the DW model (95\% CL).

Upper mass constraints from the lack of X-ray line flux are
complementary to lower mass constraints from observed cosmological
structure.  Since sterile neutrinos behave as increasingly warm dark
matter for lighter particle masses, observations of the lack of
deviations from absolute cold dark matter clustering constrain lighter
particle masses.  Two recent analyses of a single measurement of the
flux power spectrum of the Ly$\alpha$ forest from the Sloan Digital
Sky Survey (SDSS)~\cite{McDonald:2004eu} find roughly similar
constraints $m_s > 14\rm\ keV$~\cite{Seljak:2006qw} and $m_s \gtrsim
9\rm\ keV$ \cite{Viel:2006kd}\footnote{We have applied a rough 10\%
  reduction in the limit of Ref.~\cite{Viel:2006kd} in order to
  include relevant nonthermal effects, though this correction is
  significantly mass dependent~\cite{AbazajianProduction05}.} in the
DW model.  These improved considerably relative to previous
constraints at $m_s \gtrsim 2\rm\ keV$, which used the inferred linear
matter power spectrum from the SDSS and higher-resolution Ly$\alpha$
flux power spectra \cite{Viel:2005qj,Abazajian:2005xn}.  The
improvement by a factor of five stems from the high-redshift ($z\sim
4$) Ly$\alpha$ flux power spectra of the SDSS, where there is less
enhancement of the amplitude of matter power at small scales due to
the nonlinear growth of structure.  The newer Ly$\alpha$ limits are
quite stringent.  When combined with X-ray observations of Andromeda
and other mass halos, they would exclude the DW production mechanism
for sterile neutrinos.  However, it should be noted that the amplitude
and slope of the dark matter power spectrum inferred from the flux
power spectrum measurement of Ref.~\cite{McDonald:2004eu} used in both
sterile neutrino analyses is inconsistent with that inferred from the
Wilkinson Microwave Anisotropy Probe (WMAP) third year
data~\cite{Spergel:2006hy} and also indicates a number of relativistic
degrees of freedom of $N_\nu= 5.4^{+0.4}_{-0.6}$~\cite{Seljak:2006bg},
which is in tension both with that expected $N_\nu=
3.046$~\cite{Mangano:2005cc} for the case of active neutrinos alone
and that constrained by primordial nucleosynthesis, $N_\nu=
3.08^{+0.74}_{-0.68}$ \cite{Cyburt:2004yc}. This may indicate hidden
systematic effects within the measurement presented in
Ref.~\cite{McDonald:2004eu} of the Ly$\alpha$ flux power spectrum
which could alleviate or remove constraints on the inferred matter
power spectrum from the Ly$\alpha$ forest~\cite{Hui}.

In this paper, we use \chandra\ spectra of the unresolved component of
the Cosmic X-ray Background (CXB) in the direction of \chandra\ Deep
Fields North and South (hereafter CDFN and CDFS)
\cite{Hickox:2005dz,HickoxInPrep} to search for the contribution from
the radiative decay of a sterile neutrino which could comprise the
dark matter halo of the Milky Way (MW). In addition, we consider the
sensitivity of an existing short measurement at lower energies (0.4--1
keV) with an X-ray calorimeter \cite{McCammon:2002gb} to the sterile
neutrino signal.

%
\section{The Dark Matter Model}
\label{sec:dmmodel}

For a sterile neutrino of mass $m_s$ and a mixing angle $\theta$, the
decay rate for a Dirac-type active-sterile mass coupling is given by
\cite{Pal:1981rm,Barger:1995ty}
\begin{equation}
  \Gamma_\gamma(m_s,\theta) = 1.36\times 10^{-29}{\rm\ s}^{-1}\ 
  \left(\frac{\sin^2 2\theta}{10^{-7}}\right)
  \left(\frac{m_s}{1\rm\ keV}\right)^5.
\label{eq:rate}
\end{equation}
Note that here we identify the particle mass $m_s$ with that of the
mass eigenstate most closely associated with the sterile neutrino.
The dominantly sterile neutrino mass eigenstate decays into a photon
of $E=m_s/2$ and a predominantly active neutrino mass eigenstate.  The
decay rate as a function of particle mass, for a ratio of dark matter
density density to critical density $\Omega_{\rm s}$ and a fixed
quark-hadron transition temperature is given by
\begin{equation}
  \Gamma_\gamma^{\rm DW} (m_s) = 9.95\times 10^{-30}{\rm\ s}^{-1}\
  \left(\frac{m_s}{1\rm\ keV}\right)^{3.37}
  \left(\frac{\Omega_{\rm s}}{0.26}\right).
  \label{eq:rateDW}
\end{equation}
Here, we use the inferred relation between mass and mixing angle which
is necessary to produce the appropriate cosmological density of dark
matter in sterile neutrinos through the DW
mechanism~\cite{AbazajianProduction05}. We assume a cross-over
transition for the quark-hadron at a temperature of 170 MeV for the
relatoin, Eq.~(\ref{eq:rateDW}).  In general, this can vary
significantly due to uncertainties in the nature of the quark-hadron
transition, as pointed out in Abazajian \& Fuller
\cite{Abazajian:2002yz} and shown in greater detail recently by Asaka,
Laine \& Shaposhnikov \cite{Asaka:2006nq}.

The expected flux from the decay of sterile neutrino dark matter
depends on the mass of the particle, its decay rate, as well as the
distance and distribution of dark matter across the field of view of a
detector, such as {\em Chandra}.  Estimates of the dark matter mass
within the field of view of the detector come from dynamic measures of
the galactic mass profiles, or for clusters of galaxies, from the
X-ray hydrostatic method or gravitational lensing.

The flux in the neutrino decay line reaching the detector is
calculated as follows. Denoting $x$\/ the linear coordinate along the
line of sight and $d\Omega$ an element of the detector field of view
(FOV), each corresponding volume element $x^2\,d\Omega\,dx$ contains
$\rho_{\rm dm}(x)/m_s$ neutrinos, each decaying with a frequency
$\Gamma_\gamma$ given by Eqs.\ (\ref{eq:rate}--\ref{eq:rateDW}). In
Euclidean geometry (i.e., as long as the object redshift $z$ satisfies
$(1+z)^4\approx 1$), the photon flux from this volume element reaching
a unit effective area of the detector is
\begin{equation}
df = x^2\,d\Omega\,dx \frac{\Gamma_\gamma\,\rho_{\rm dm}(x)}{m_s} \frac{1}{4\pi x^2}.
\end{equation}
The flux from the cone subtended by the detector FOV (in
photons\,s$^{-1}$\,cm$^{-2}$) is therefore
\begin{equation}
f = \frac{\Gamma_\gamma}{4\pi\,m_s} \int d\Omega \int \rho_{\rm dm}(x)\,dx
  = \frac{\Gamma_\gamma}{4\pi\,m_s} \Omega \bar{S}_{\rm dm},
\label{eq:f}
\end{equation}
where the first integral is over the FOV and the second integral is
along the line of sight, and $\bar{S}_{\rm dm}$ is the dark matter
mass column density averaged over the FOV $\Omega$.  The flux coming
from each direction in the FOV depends only on $S_{\rm dm}$ in that
direction; details of the line of sight mass distribution, and even
the distance to the object, are not important for objects at low $z$.
For \chandra's small FOV (a $r=5'$ circle used in this work), we can
ignore the changes of the Milky Way mass column density across the
FOV.  However, in \S\ref{calorimetry} we will consider an instrument
with a much wider FOV, for which we directly integrate the column
density within the FOV.

To calculate $S_{\rm dm}$ for the Milky Way halo in the directions of
the two \chandra\ CXB observations, we use mass models from Galactic
dynamics measures presented in Klypin et al.~\cite{Klypin2001} and
Battaglia et al.~\cite{Battaglia:2005rj}, which are both consistent
with each other and with Ref.~\cite{Widrow:2005bt}.  We use the
Navarro-Frenk-White (NFW) profile~\cite{NFW96}
\begin{equation}
\rho(r)=\rho_s \left(\frac{r}{r_s}\right)^{-1} 
               \left(1+\frac{r}{r_s}\right)^{-2},
\end{equation}
where $\rho_s$ is the characteristic density and $r_s$ is the scale
radius.  The NFW profile is expected for CDM and should be consistent
with WDM models where cores are expected to be much smaller than the
MW NFW scale radius~\cite{Avila-Reese:2000hg}.

The dynamical constraints on the dark matter halo of the MW are
complicated by the fact that we sit within this galaxy, and depend
strongly on the measured orbital velocities of satellite dwarf
galaxies.  The 68\% confidence region of the NFW models from Battaglia
et al.\ also covers the allowed model range from Klypin et al., while
still being consistent with observations. We thus adopt that interval
for our estimates.  These models have $M_{\rm vir} = 0.6\times
10^{12}\ \msun$ (low mass MW) and $M_{\rm vir} = 2.0\times 10^{12}\
\msun$ (high mass MW), $R_{\rm vir} = 255\rm\ kpc$, $R_{\rm
  vir}/r_s=18$, and $R_\odot=8$ kpc.  In both models, most of the
baryonic mass is concentrated in the central region of the MW (disk
and bulge), which our lines of sight miss, thus we can simply (and
somewhat conservatively) multiply the total density with the universal
dark matter fraction $f_{\rm DM} = \Omega_{\rm DM}/(\Omega_{\rm
  DM}+\Omega_b) \approx 0.867$, where the fraction of critical density
of the dark matter we take is $\Omega_{\rm DM} = 0.26$, and baryon
density $\Omega_b = 0.04$.  The uncertainties in the MW halo model are
much larger than that of the universal dark matter fraction, which we
keep fixed.

For a point at a distance $x$ from the Sun along the line of sight with
Galactic coordinates $(\ell,b)$, the distance from the center of the halo is
\begin{equation}
r=(x^2 - 2 x R_\odot \cos b \cos \ell + R_\odot^2)^{1/2},
\end{equation}
which is used to evaluate the column density integral in
Eq.~(\ref{eq:f}).  In the direction of the CDFN at
$\ell=125.89^\circ$, $b=54.83^\circ$, the dark matter surface density
is
\begin{equation}
S_{\rm dm} = 
\begin{cases}
0.0362\rm\ g\ cm^{-2}\quad\text{(high mass MW)}\\
0.0109\rm\ g\ cm^{-2}\quad\text{(low mass MW)}.
\end{cases}
\end{equation}
For the CDFS at $\ell=223.57^\circ$, $b=-54.44^\circ$, the
corresponding surface densities are 4\% lower.

We integrated the MW halo out to $R_{\rm vir}$. However, most of the
column density accumulates at small radii (within $0.1 R_{\rm vir}$),
so the outer radial cutoff does not matter. It also means that the
column density depends on small-scale deviations from the symmetric
model. Therefore, our range of dark matter column densities is only
qualitative and does not reflect the full uncertainty, which is
difficult to quantify, and therefore we leave our inferred limits with
as they are but with the caveat of potentially non-negligible
systematic uncertainty.  A substantive measure of the systematic
uncertainty may be estimated via comparisons with other independent
measurements.

As a comparison, the mass surface density within 5\deg\ of the
Galactic center is an order of magnitude higher, as is the surface
density on the line of sight near the center of M31. Massive galaxy
clusters have column densities of $0.1-0.3$ \gcmsq\ in their core
regions.  Clearly, our lines of sight are not optimal for the neutrino
line search, and more sensitive limits can be obtained, for example,
from observations of the Galactic Center (which we will attempt in a
future work).  At the same time, our CDF datasets have the advantage
of an almost complete removal of the sources causing the CXB and a
very accurate background modeling.

\section{\chandra\ CXB spectrum}
\label{observations}

\begin{figure}[t]
\includegraphics[width=3.3truein]{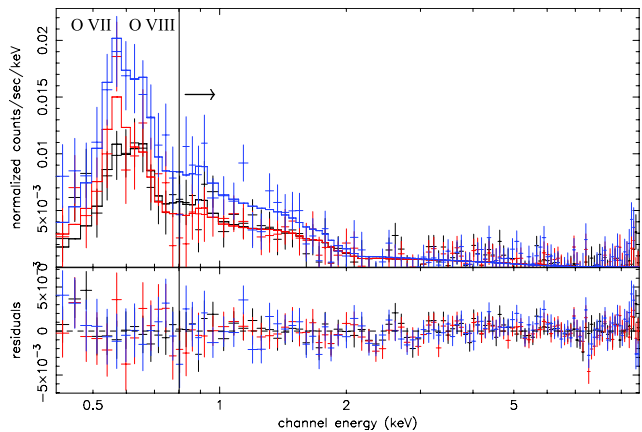}
\caption{\small The three unresolved CXB spectra [CDFN-VF (black),
  CDFN-F (red), and CDFS (blue)] in the 0.4--10 keV band. They are fit
  well by a simple model consisting of a power law absorbed by the
  Galactic hydrogen column representing the extragalactic component,
  plus unabsorbed local warm thermal emission.  All three spectra are
  fit well, with a total reduced $\chi^2$ of 0.87, and no residual
  line-like features are seen. Only energies $E> 0.8\rm\ keV$ are
  used for the line flux limits. See text for details.} 
\label{neut_spec}
\end{figure}

The {\em Chandra X-ray Observatory} \cite{Weisskopf:2001uu} has a 1
arcsecond on-axis angular resolution and is uniquely suited for
resolving the CXB covered by its ACIS detector, which operates in the
0.4--10 keV energy band.  It has performed a series of very deep
observations of two fields, CDFN and CDFS, for a total exposure of 2
Ms and 1 Ms, respectively, aimed at resolving as much of the
extragalactic CXB into point sources as possible
\cite{Brandt:2001vb,Giacconi:2001tb}. Hickox \& Markevitch
\cite{Hickox:2005dz} (hereafter H06) have used these observations,
along with recent accurate calibration of the ACIS internal
background, to derive a spectrum of the CXB that is still unresolved
after spatially excising all X-ray sources detectable in these deep
exposures. They found that about 20\% of the CXB in the 1--2 and 2--8
keV bands remains unresolved, while at $E\lesssim 1$ keV, the Galaxy
contributes a dominant genuinely diffuse component.  The 1--8 keV
unresolved spectrum is well modeled by a power law with a photon index
$\Gamma\approx 1.5$, and the Galactic diffuse component by thermal
emission from a $T\approx 0.2$ keV plasma.

These unresolved spectra do not exhibit any emission lines that can be
attributed to decaying sterile neutrinos in the Galactic halo, and we
will use this fact to place upper limits on the neutrino line in the
$E=0.8-8$ keV range.  We note that in a later work, Hickox \&
Markevitch \cite{HickoxInPrep} have additionally excised sources not
detected in X-rays but seen by the {\Hubble Space Telescope}\/ in the
optical. These sources collectively account for most of the unresolved
1--8 keV flux, with a residual CXB brightness consistent with zero.
Although this seems advantageous for our upper limits, we will not use
these new spectra, because excising those optical sources reduced the
solid angle from which the CXB spectra are collected by a factor of 3,
compared to that in H06.  This increased statistical uncertainty and
removed any advantage for our analysis.  The CXB in either case is a
small fraction of the detector background, and our constraints on the
neutrino line are limited by photon statistics of this background. We
thus elected to tolerate some unresolved CXB flux (modeled with a
power law as described below) but have a significantly smaller
statistical scatter in each spectral bin. At the same time, we will
take advantage of a longer calibration exposure of the ACIS internal
background that was used in the latter work.

Full account of the CDF spectra derivation is given in H06, and here
we give only the relevant details. The CFDN dataset is divided into
two, one observed with ACIS in FAINT mode and another in VFAINT (the
latter more advantageous for background modeling). We treat these
subsets as independent observations (CDFN-F and CDFN-VF). After
excluding periods of elevated detector background, their exposures are
472 ks and 537 ks, respectively.  The CDFS was observed in FAINT mode
and has a clean exposure of 568 ks.  Solid angles, after the source
exclusion, subtended by the CDFN and CDFS fields are 0.0135 deg$^2$
and 0.0159 deg$^2$, respectively.  The internal detector background is
modeled and subtracted using a set of calibration observations in
which ACIS was shielded from the sky.  We will use a more recent 325
ks calibration dataset from Hickox \& Markevitch \cite{HickoxInPrep} ,
compared to 236 ks in H06. This exposure is still shorter than any of
the CDF exposures, and the statistical uncertainty of this calibration
spectrum will be the main limiting factor for our analysis.
Otherwise, the CXB spectra that we analyze here are identical to those
in H06.

As noted in H06, the three unresolved CXB spectra (CDFN-VF, CDFN-F,
and CDFS) in the 0.4--10 keV band are fit well by a simple model
consisting of a power law absorbed by the Galactic hydrogen column
($N_H=1.5\times 10^{20}$\cmsq\ for CDFN and $0.9\times 10^{20}$\cmsq\
for CDFS) representing the extragalactic component, plus unabsorbed
local warm thermal emission with solar heavy element abundances (the
APEC model \cite{Smith:2001he}).  We assumed the power law to be the
same between the CDFN-VF and CDFN-F spectra, but allowed it to be
different for CDFS, since it points to a different region of the
sky. Thermal models were allowed to be different between all three
datasets, because they may also include a time-variable near-Earth
charge exchange contribution (the dominant features of both the
thermal and charge-exchange components is an O{\sc vii} line around
570 eV, so they are difficult to disentangle).

The resulting fits are shown in Fig.~\ref{neut_spec}. All three
spectra are fit well, with a total reduced $\chi^2$ of 0.87; no
residual line-like features are seen. We can place an upper limit on
such a line as a function of energy, excluding the range $E<0.8$ keV,
where such a limit would not be interesting because of strong emission
lines in the Galactic thermal spectrum, not resolved well by the ACIS
CCD. To do this, we added a monochromatic line to the spectral model
with the relative normalizations between the CDF regions determined
from the predicted neutrino brightness in the directions of those
fields (\S\ref{sec:dmmodel}) and the region solid angles. For both the
low-mass and high-mass Galaxy models, this ratio for CDFS/CDFN is
1.14. For each line energy in the range 0.8--9 keV, we re-fit the
model and derived upper limits on the neutrino line normalization,
allowing all other model parameters to vary as above (so that the
neutrino line was allowed to account for the flux from other model
components).

\begin{figure}[t]
\includegraphics[width=3.3truein]{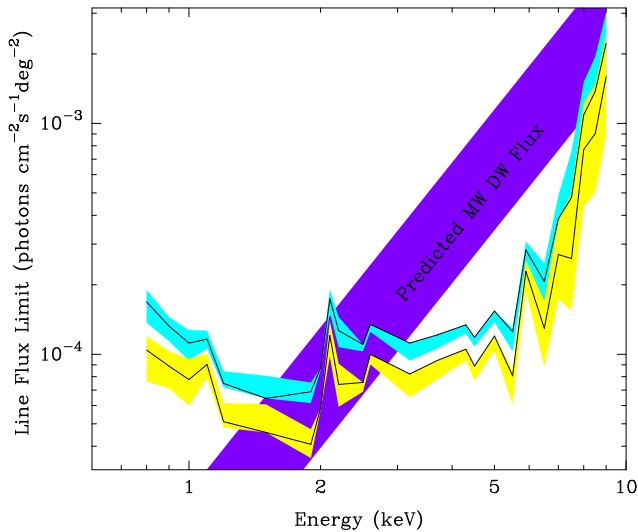}
\caption{\small Shown are the limits on the flux in a line as a
  function of energy from the CDFN-VF, CDFN-F, and CDFS.  For each
  line energy in the range 0.8--9 keV, all model parameters were
  allowed to vary while deriving upper limits on the sterile neutrino
  line normalization.  The upper line and (cyan) band are the
  $3\sigma$ limit, with the band representing uncertainty in the
  detector background modeling.  The lower band and line correspond to
  that from the $2\sigma$ limits.  The diagonal band is the range of
  expected line flux from the MW halo as a function of energy for
  the sterile neutrino in a DW production model. }
\label{neut_lineflux}
\end{figure}

\begin{figure}[t]
\includegraphics[width=3.3truein]{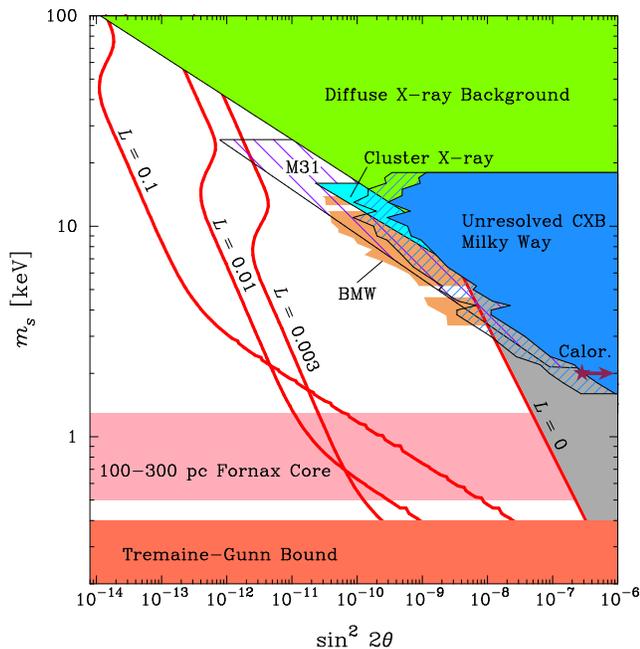}
\caption{\small Full parameter space constraints for the sterile
  neutrino production models, assuming sterile neutrinos constitute
  the dark matter.  Contours labeled with lepton number $L=0$,
  $L=0.003$, $L=0.01$, $L=0.1$ are production predictions for constant
  comoving density of $\Omega_s=0.24$ for $L=0$, and $\Omega_s=0.3$
  for non-zero $L$~\cite{Abazajian:2002yz}.  Constraints from the CXB
  with the minimal MW halo model are in the solid (blue) region, while
  the maximal MW halo model excludes the adjacent diagonally hatched
  region. Also shown are exclusion regions from the diffuse X-ray
  background (green)~\cite{Boyarsky:2005us}, from XMM-Newton
  observations of the Coma and Virgo clusters (light
  blue)~\cite{Boyarsky:2006zi}, observations of Andromeda (M31) in
  wide hatching~\cite{Watson:2006qb}, and limits from the MW by
  Boyarsky et al.~\cite{Boyarsky:2006ag} (BMW).  The region at
  $m_s<0.4\rm\ keV$ is ruled out by a conservative application of the
  Tremaine-Gunn bound~\cite{Bode:2000gq}.  The grey region to the
  right of the $L=0$ case is where sterile neutrino dark matter is
  over-produced.  The constraint from the MW calorimeter soft X-ray
  background observation is the star and arrow, marked ``Calor.'' Also
  shown is the horizontal band of the mass scale consistent with
  producing a 100--300 pc core in the Fornax dwarf
  galaxy~\cite{Strigari:2006ue}. The non-resonant and resonant
  production curves come from Refs.~\cite{AbazajianProduction05} and
  \cite{Abazajian:2002yz}, respectively.}
\label{omega_full}
\end{figure}

The spectra were binned in such a way that an emission line would be
resolved at all energies and the statistics in each bin is Gaussian.
The upper limits on the line normalization can then be derived from
$\Delta \chi^2$ w.r.t.\ the best-fit model. It is an adequate estimate
(cf.\ Ref.~\cite{Protassov:2002sz}), because the sky signal is a small
fraction of the underlying detector background count rate (20\% and
3\% in the 1--2 and 2--8 keV bands, respectively), which dominates the
statistics.  We note that while the CDF spectra are statistically
independent, the detector background spectrum is almost the same for
all three datasets, which introduces correlations among the three
resulting CXB spectra.  This has been taken into account by a simple
Monte-Carlo simulation, from which we determined that for our
particular combination of the CDF and background exposures, $2\sigma$
and $3\sigma$ upper limits for one interesting parameter correspond to
$\Delta \chi^2$ of 8.8 and 19.7, respectively (compared to 4 and 9 for
uncorrelated spectra). Using these values, we obtained upper limits on
the neutrino line flux shown in Fig.\ \ref{neut_lineflux}. The
constraints generally follow the energy dependence of the ACIS
effective area and the presence of bright background lines (e.g., the
fluorescent Au line at $E=2.1$ keV). In addition, we varied the
normalization of the detector background spectrum by $\pm3$\%
(simultaneously for all three spectra), which represents the
background modeling uncertainty (H06). The effect of this is shown as
bands around the 2 and 3$\sigma$ limits in Fig.~\ref{neut_lineflux}.
We use the upper envelopes of these bands for the constraints below.
The parameter space excluded by these constraints are shown, relative
to other constraints, in Fig.~\ref{omega_full}.

In the DW model, these upper limits
correspond to $2\sigma$ limits on the sterile neutrino
particle mass
\begin{equation}
m_s <
\begin{cases} 
  2.87 {\rm\ keV}\quad \text{(high mass MW)}\\
  5.66 {\rm\ keV}\quad \text{(low mass MW)}.
\end{cases}
\end{equation}
As is clear from these limits, there is considerable uncertainty due
to the modeling of the dynamics of the MW halo.

\section{Observations of the Soft X-ray Background}
\label{calorimetry}

\begin{figure}[t]
\includegraphics[width=3.3truein]{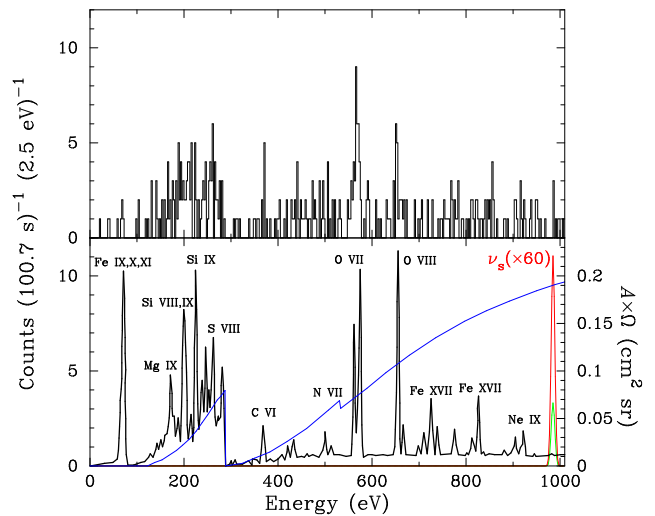}
\caption{\small The upper panel is the spectrum soft X-ray background
  as measured by McCammon et al.~\cite{McCammon:2002gb}.  The lower
  panel shows the atomic line model, detector response function (blue)
  and expected contribution ($\times$60) due to sterile neutrino decay
  of the DW production model with the minimal (lower, green) and
  maximal (upper, red) MW halo models.}
\label{soft_xrb}
\end{figure}

Here we explore the constraints from a related observation at $E<1$
keV using a detector with a much higher spectral resolution.  The
observation was made with an X-ray calorimeter flown on a sounding
rocket, as reported by McCammon et al.~\cite{McCammon:2002gb}. The
spectrum from their 100 s exposure is shown in the upper panel of
Fig.~\ref{soft_xrb}. The observation was made towards Galactic
coordinates $\ell=90^\circ, b=+60^\circ$. The mass surface density
averaged over the wide ($\Omega=0.81$ sr) field of view of this
experiment is
\begin{equation}
\bar{S}_{\rm dm} = 
\begin{cases}
0.0460\rm\ g\ cm^{-2}\quad\text{(high mass MW)}\\
0.0138\rm\ g\ cm^{-2}\quad\text{(low mass MW)}.
\end{cases}
\end{equation}
The quantity $\Omega \bar{S}_{\rm dm}$, which determines the expected
sterile neutrino decay signal in the spectrum, is $2.5\times 10^{5}$
times higher than in the CDF observations. The dwarf spheroidal
galaxies Ursa Minor and Draco lie within the FOV of this observation,
but contribute only 2\% to $\bar{S}_{\rm dm}$.

To get a rough idea of how sensitive this experiment is to the sterile
neutrino decay, we estimated the neutrino line flux at $E=1$
keV. Using the detector response and exposure time from McCammon et
al., we find that an emission line from the $m_s=2$ keV sterile
neutrino in the DW model would contain 1.8 or 0.5 counts in our
maximum and minimum halo mass models, respectively. This is obviously
too low to be detectable in the present data.  A rough upper limit on
the line flux at $E=1$ keV can be derived in the following way. The
observed background in the McCammon spectrum at 1 keV is $\sim 1.5$
counts per 2.5 eV bin, so $\sim 5$ counts within FWHM of 9 eV. For
Poissonian statistics, for an emission line on top of this background,
a $3\sigma$ upper limit is $\sim 12$ counts (9 counts within the
FWHM), or $2\sigma$ limit of 6 counts.  This would place a 2$\sigma$
limit between 3 and 11 times the mixing angle $\sin^2 2\theta$
predicted by the DW model, for the high and low mass MW models,
respectively, at a sterile neutrino particle mass of 2 keV.  For the
low mass case, this corresponds to excluding mixing angles at $\sin^2
2\theta \gtrsim 3\times 10^{-7}$ (see Fig.~\ref{omega_full}).  At
lower energies the detector efficiency and the neutrino decay rate
both decline, so the constraints weaken rapidly. Nevertheless, the
above limit for $m_s=2$ keV is similar to our ACIS constraints from a
much longer observation, which shows the huge benefit of a wide FOV
combined with a calorimetric energy resolution.

\section{Discussion and Conclusions}
\label{discussion}

The measurement by Hickox \& Markevitch
\cite{Hickox:2005dz,HickoxInPrep} of the unresolved cosmic X-ray
background, produced by a total of 3 Ms observations by the ACIS CCD
aboard the \chandra\ X-ray telescope, presents a constraint on the
presence of a line flux that would be expected due to the decay of the
sterile neutrino dark matter candidate comprising the dark matter halo
of the MW.  We find that the $2\sigma$ upper limits on the sterile
neutrino particle mass in the simplest (DW) production mechanism are
in the range of $m_s < 2.9$ and $m_s < 5.7$ keV for high and low mass
estimates of the MW dark matter halo, respectively.  To be
conservative, the higher value for the upper limit of 5.7 keV is the
robust 95\%~CL upper limit to conclude from the CXB analysis in the
work presented here.  More accurate dynamical measures of the MW halo
are directly relevant for the particle mass constraints presented
here.  In addition, we find that there is a significant limit to
potential constraints from the \chandra\ observatory at photon
energies $E_\gamma \lesssim 1\rm\ keV$ due to the presence of emmision
lines of the low temperature gas in the MW as well as the rapidly
worsening relative energy resolution of the ACIS (as well as XMM EPIC)
detector at these energies. This limits potential constraints from
\chandra\ (or {\it XMM-Newton}) on the sterile neutrino particle mass
to not better than $m_s \approx 2\rm\ keV$.

The limits presented here from the MW contribution to the measured CXB
in \chandra\ are comparable to that from the {\it XMM-Newton}
observation of Andromeda, $m_s < 3.5\rm\ keV$ (95\% CL) for the DW
model, by Watson et al.~\cite{Watson:2006qb} who use a stringent
requirement of the decay signal to be four times the astrophysical
background within a bin to place this upper limit.  Their
constraints in the full parameter space (Fig.~\ref{omega_full})
were derived using an energy-averaged flux-to-counts conversion
instead of the direct spectral fitting, which resulted in an
approximate power-law constraint. In reality, their constraints must
weaken at high energies as they do in this work, because the {\it
  XMM-Newton} effective area also declines sharply with energy.

The MW constraints by Boyarsky et al.~\cite{Boyarsky:2006fg} from the
{\it XMM-Newton} X-ray background data in the full parameter space are
comparable to ours.  Their more recent analysis of the XMM background
data~\cite{Boyarsky:2006ag} resulted in $3\sigma$ constraints plotted
in Fig.\ 3 (where the horizontal gaps correspond to the intervals
$E=2.3-2.6$ keV and $5.9-6.3$ keV excluded from their spectral
fits). Their adopted MW halo model lies between our minimum and
maximum mass models.  While \chandra\ has an advantage over {\it
  XMM-Newton} of a more stable instrumental background and almost
complete removal of the CXB in our deep exposures, the solid angle
subtended by the {\it XMM-Newton} FOV is about 10 times greater than
ours. This results in a proportionally higher expected MW sterile
neutrino signal and more stringent constraints.  On the other hand,
because of the above two reasons, we could afford to derive
constraints in a slightly wider energy range.

Reimer-Sorenson et al.~\cite{Riemer-Sorensen:2006fh} also analyzed
\chandra\ CXB observations using the blank-sky data, and derived
constraints on the MW sterile neutrino flux. Because they did not
remove the ACIS internal background (that dominates at all energies
and is well calibrated), their constraints on the flux are about two
orders of magnitude weaker than ours.

We have also analyzed a related observation by McCammon et
al.~\cite{McCammon:2002gb} in the soft X-ray, motivated by the high
spectral resolution of the observation at low photon energies, as well
as the large field of view (0.81 ster) providing a corresponding two
orders of magnitude larger MW halo mass in the field of view compared
to the CDF observations.  We find that though this observation in
itself does not provide competitive constraints in the DW production
model, longer exposure observations with low backgrounds may detect or
place constraints within a photon energy range ($E_\gamma < 1\rm\
keV$) and particle mass range ($m_s < 2\rm\ keV$), which is
inaccessible to current X-ray telescopes.  This observation constrains
$m_s = 2$ keV neutrinos to not have $\sin^22\theta \gtrsim 3\times
10^{-7}$. Although this region is excluded in the DW model due to
over-production, the constraint may be of interest for low reheating
temperature or dilution production
models~\cite{Asaka:2006ek,Gelmini:2004ah}.  Future observations
of the type by McCammon et al.\ in the soft X-ray, as well as those
possible with the {\it Constellation-X} observatory at higher photon
energies~\cite{Abazajian:2001vt}, may probe the full parameter space
for oscillation-based sterile neutrino dark matter production models.

\acknowledgments

We would like to thank John Bea\-com, Dan\-iel Boy\-a\-novsky, A.\
Boyarsky, Ge\-org Raf\-felt, Oleg Ruchayskiy and Louie Strigari for
useful discussions.  KNA would like to thank the Harvard-Smithsonian
Center for Astrophysics for hospitality, where initial work on this
project took place, and the Max Planck Institute for Physics in Munich
for hospitality, where it was completed.  MM was supported by NASA
contract NAS8-39073 and \chandra\ grant GO4-5152X.  Work performed at
LANL was carried out under the auspices of the NNSA of the U.\ S.\
Department of Energy at Los Alamos National Laboratory under Contract
No. DE-AC52-06NA25396.

\bibliography{master.bib}

\begin{thebibliography}{68}
\expandafter\ifx\csname natexlab\endcsname\relax\def\natexlab#1{#1}\fi
\expandafter\ifx\csname bibnamefont\endcsname\relax
  \def\bibnamefont#1{#1}\fi
\expandafter\ifx\csname bibfnamefont\endcsname\relax
  \def\bibfnamefont#1{#1}\fi
\expandafter\ifx\csname citenamefont\endcsname\relax
  \def\citenamefont#1{#1}\fi
\expandafter\ifx\csname url\endcsname\relax
  \def\url#1{\texttt{#1}}\fi
\expandafter\ifx\csname urlprefix\endcsname\relax\def\urlprefix{URL }\fi
\providecommand{\bibinfo}[2]{#2}
\providecommand{\eprint}[2][]{\url{#2}}

\bibitem[{\citenamefont{Spergel et~al.}(2006)}]{Spergel:2006hy}
\bibinfo{author}{\bibfnamefont{D.~N.} \bibnamefont{Spergel}}
  \bibnamefont{et~al.}, \bibinfo{journal}{submitted, Astrophys. J.}
  (\bibinfo{year}{2006}), \eprint{astro-ph/0603449}.

\bibitem[{\citenamefont{Tegmark et~al.}(2006)}]{Tegmark:2006az}
\bibinfo{author}{\bibfnamefont{M.}~\bibnamefont{Tegmark}} \bibnamefont{et~al.},
  \bibinfo{journal}{Phys. Rev.} \textbf{\bibinfo{volume}{D74}},
  \bibinfo{pages}{123507} (\bibinfo{year}{2006}), \eprint{astro-ph/0608632}.

\bibitem[{\citenamefont{Goerdt et~al.}(2006)\citenamefont{Goerdt, Moore, Read,
  Stadel, and Zemp}}]{Goerdt:2006rw}
\bibinfo{author}{\bibfnamefont{T.}~\bibnamefont{Goerdt}},
  \bibinfo{author}{\bibfnamefont{B.}~\bibnamefont{Moore}},
  \bibinfo{author}{\bibfnamefont{J.~I.} \bibnamefont{Read}},
  \bibinfo{author}{\bibfnamefont{J.}~\bibnamefont{Stadel}}, \bibnamefont{and}
  \bibinfo{author}{\bibfnamefont{M.}~\bibnamefont{Zemp}},
  \bibinfo{journal}{Mon. Not. Roy. Astron. Soc.}
  \textbf{\bibinfo{volume}{368}}, \bibinfo{pages}{1073} (\bibinfo{year}{2006}),
  \eprint{astro-ph/0601404}.

\bibitem[{\citenamefont{Sanchez-Salcedo
  et~al.}(2006)\citenamefont{Sanchez-Salcedo, Reyes-Iturbide, and
  Hernandez}}]{Sanchez-Salcedo:2006fa}
\bibinfo{author}{\bibfnamefont{F.~J.} \bibnamefont{Sanchez-Salcedo}},
  \bibinfo{author}{\bibfnamefont{J.}~\bibnamefont{Reyes-Iturbide}},
  \bibnamefont{and}
  \bibinfo{author}{\bibfnamefont{X.}~\bibnamefont{Hernandez}},
  \bibinfo{journal}{Mon. Not. Roy. Astron. Soc.}
  \textbf{\bibinfo{volume}{370}}, \bibinfo{pages}{1829} (\bibinfo{year}{2006}),
  \eprint{astro-ph/0601490}.

\bibitem[{\citenamefont{Strigari et~al.}(2006{\natexlab{a}})}]{Strigari:2006ue}
\bibinfo{author}{\bibfnamefont{L.~E.} \bibnamefont{Strigari}}
  \bibnamefont{et~al.}, \bibinfo{journal}{Astrophys. J.}
  \textbf{\bibinfo{volume}{652}}, \bibinfo{pages}{306}
  (\bibinfo{year}{2006}{\natexlab{a}}), \eprint{astro-ph/0603775}.

\bibitem[{\citenamefont{Gilmore et~al.}(2006)}]{Gilmore:2006iy}
\bibinfo{author}{\bibfnamefont{G.}~\bibnamefont{Gilmore}} \bibnamefont{et~al.}
  (\bibinfo{year}{2006}), \eprint{astro-ph/0608528}.

\bibitem[{\citenamefont{Klypin et~al.}(1999)\citenamefont{Klypin, Kravtsov,
  Valenzuela, and Prada}}]{Klypin:1999uc}
\bibinfo{author}{\bibfnamefont{A.~A.} \bibnamefont{Klypin}},
  \bibinfo{author}{\bibfnamefont{A.~V.} \bibnamefont{Kravtsov}},
  \bibinfo{author}{\bibfnamefont{O.}~\bibnamefont{Valenzuela}},
  \bibnamefont{and} \bibinfo{author}{\bibfnamefont{F.}~\bibnamefont{Prada}},
  \bibinfo{journal}{Astrophys. J.} \textbf{\bibinfo{volume}{522}},
  \bibinfo{pages}{82} (\bibinfo{year}{1999}), \eprint{astro-ph/9901240}.

\bibitem[{\citenamefont{Moore et~al.}(1999)}]{Moore:1999wf}
\bibinfo{author}{\bibfnamefont{B.}~\bibnamefont{Moore}} \bibnamefont{et~al.},
  \bibinfo{journal}{Astrophys. J.} \textbf{\bibinfo{volume}{524}},
  \bibinfo{pages}{L19} (\bibinfo{year}{1999}), \eprint{astro-ph/9907411}.

\bibitem[{\citenamefont{Blumenthal et~al.}(1982)\citenamefont{Blumenthal,
  Pagels, and Primack}}]{Blumenthal:1982mv}
\bibinfo{author}{\bibfnamefont{G.~R.} \bibnamefont{Blumenthal}},
  \bibinfo{author}{\bibfnamefont{H.}~\bibnamefont{Pagels}}, \bibnamefont{and}
  \bibinfo{author}{\bibfnamefont{J.~R.} \bibnamefont{Primack}},
  \bibinfo{journal}{Nature} \textbf{\bibinfo{volume}{299}}, \bibinfo{pages}{37}
  (\bibinfo{year}{1982}).

\bibitem[{\citenamefont{Bode et~al.}(2001)\citenamefont{Bode, Ostriker, and
  Turok}}]{Bode:2000gq}
\bibinfo{author}{\bibfnamefont{P.}~\bibnamefont{Bode}},
  \bibinfo{author}{\bibfnamefont{J.~P.} \bibnamefont{Ostriker}},
  \bibnamefont{and} \bibinfo{author}{\bibfnamefont{N.}~\bibnamefont{Turok}},
  \bibinfo{journal}{Astrophys. J.} \textbf{\bibinfo{volume}{556}},
  \bibinfo{pages}{93} (\bibinfo{year}{2001}), \eprint{astro-ph/0010389}.

\bibitem[{\citenamefont{Kaplinghat}(2005)}]{Kaplinghat:2005sy}
\bibinfo{author}{\bibfnamefont{M.}~\bibnamefont{Kaplinghat}},
  \bibinfo{journal}{Phys. Rev.} \textbf{\bibinfo{volume}{D72}},
  \bibinfo{pages}{063510} (\bibinfo{year}{2005}), \eprint{astro-ph/0507300}.

\bibitem[{\citenamefont{Cembranos et~al.}(2005)\citenamefont{Cembranos, Feng,
  Rajaraman, and Takayama}}]{Cembranos:2005us}
\bibinfo{author}{\bibfnamefont{J.~A.~R.} \bibnamefont{Cembranos}},
  \bibinfo{author}{\bibfnamefont{J.~L.} \bibnamefont{Feng}},
  \bibinfo{author}{\bibfnamefont{A.}~\bibnamefont{Rajaraman}},
  \bibnamefont{and} \bibinfo{author}{\bibfnamefont{F.}~\bibnamefont{Takayama}},
  \bibinfo{journal}{Phys. Rev. Lett.} \textbf{\bibinfo{volume}{95}},
  \bibinfo{pages}{181301} (\bibinfo{year}{2005}), \eprint{hep-ph/0507150}.

\bibitem[{\citenamefont{Hu et~al.}(2000)\citenamefont{Hu, Barkana, and
  Gruzinov}}]{Hu:2000ke}
\bibinfo{author}{\bibfnamefont{W.}~\bibnamefont{Hu}},
  \bibinfo{author}{\bibfnamefont{R.}~\bibnamefont{Barkana}}, \bibnamefont{and}
  \bibinfo{author}{\bibfnamefont{A.}~\bibnamefont{Gruzinov}},
  \bibinfo{journal}{Phys. Rev. Lett.} \textbf{\bibinfo{volume}{85}},
  \bibinfo{pages}{1158} (\bibinfo{year}{2000}), \eprint{astro-ph/0003365}.

\bibitem[{\citenamefont{Strigari
  et~al.}(2006{\natexlab{b}})\citenamefont{Strigari, Kaplinghat, and
  Bullock}}]{Strigari:2006jf}
\bibinfo{author}{\bibfnamefont{L.~E.} \bibnamefont{Strigari}},
  \bibinfo{author}{\bibfnamefont{M.}~\bibnamefont{Kaplinghat}},
  \bibnamefont{and} \bibinfo{author}{\bibfnamefont{J.~S.}
  \bibnamefont{Bullock}} (\bibinfo{year}{2006}{\natexlab{b}}),
  \eprint{astro-ph/0606281}.

\bibitem[{\citenamefont{Berezhiani and Mohapatra}(1995)}]{Berezhiani:1995yi}
\bibinfo{author}{\bibfnamefont{Z.~G.} \bibnamefont{Berezhiani}}
  \bibnamefont{and} \bibinfo{author}{\bibfnamefont{R.~N.}
  \bibnamefont{Mohapatra}}, \bibinfo{journal}{Phys. Rev.}
  \textbf{\bibinfo{volume}{D52}}, \bibinfo{pages}{6607} (\bibinfo{year}{1995}),
  \eprint{hep-ph/9505385}.

\bibitem[{\citenamefont{Chun and Kim}(1999)}]{Chun:1999cq}
\bibinfo{author}{\bibfnamefont{E.~J.} \bibnamefont{Chun}} \bibnamefont{and}
  \bibinfo{author}{\bibfnamefont{H.~B.} \bibnamefont{Kim}},
  \bibinfo{journal}{Phys. Rev.} \textbf{\bibinfo{volume}{D60}},
  \bibinfo{pages}{095006} (\bibinfo{year}{1999}), \eprint{hep-ph/9906392}.

\bibitem[{\citenamefont{Langacker}(1998)}]{Langacker:1998ut}
\bibinfo{author}{\bibfnamefont{P.}~\bibnamefont{Langacker}},
  \bibinfo{journal}{Phys. Rev.} \textbf{\bibinfo{volume}{D58}},
  \bibinfo{pages}{093017} (\bibinfo{year}{1998}), \eprint{hep-ph/9805281}.

\bibitem[{\citenamefont{Arkani-Hamed et~al.}(2002)\citenamefont{Arkani-Hamed,
  Dimopoulos, Dvali, and March-Russell}}]{Arkani-Hamed:1998vp}
\bibinfo{author}{\bibfnamefont{N.}~\bibnamefont{Arkani-Hamed}},
  \bibinfo{author}{\bibfnamefont{S.}~\bibnamefont{Dimopoulos}},
  \bibinfo{author}{\bibfnamefont{G.~R.} \bibnamefont{Dvali}}, \bibnamefont{and}
  \bibinfo{author}{\bibfnamefont{J.}~\bibnamefont{March-Russell}},
  \bibinfo{journal}{Phys. Rev.} \textbf{\bibinfo{volume}{D65}},
  \bibinfo{pages}{024032} (\bibinfo{year}{2002}), \eprint{hep-ph/9811448}.

\bibitem[{\citenamefont{Abazajian et~al.}(2003)\citenamefont{Abazajian, Fuller,
  and Patel}}]{Abazajian:2000hw}
\bibinfo{author}{\bibfnamefont{K.}~\bibnamefont{Abazajian}},
  \bibinfo{author}{\bibfnamefont{G.~M.} \bibnamefont{Fuller}},
  \bibnamefont{and} \bibinfo{author}{\bibfnamefont{M.}~\bibnamefont{Patel}},
  \bibinfo{journal}{Phys. Rev. Lett.} \textbf{\bibinfo{volume}{90}},
  \bibinfo{pages}{061301} (\bibinfo{year}{2003}), \eprint{hep-ph/0011048}.

\bibitem[{\citenamefont{Asaka et~al.}(2005)\citenamefont{Asaka, Blanchet, and
  Shaposhnikov}}]{Asaka:2005an}
\bibinfo{author}{\bibfnamefont{T.}~\bibnamefont{Asaka}},
  \bibinfo{author}{\bibfnamefont{S.}~\bibnamefont{Blanchet}}, \bibnamefont{and}
  \bibinfo{author}{\bibfnamefont{M.}~\bibnamefont{Shaposhnikov}},
  \bibinfo{journal}{Phys. Lett.} \textbf{\bibinfo{volume}{B631}},
  \bibinfo{pages}{151} (\bibinfo{year}{2005}), \eprint{hep-ph/0503065}.

\bibitem[{\citenamefont{Kusenko and Segre}(1999)}]{Kusenko:1998bk}
\bibinfo{author}{\bibfnamefont{A.}~\bibnamefont{Kusenko}} \bibnamefont{and}
  \bibinfo{author}{\bibfnamefont{G.}~\bibnamefont{Segre}},
  \bibinfo{journal}{Phys. Rev.} \textbf{\bibinfo{volume}{D59}},
  \bibinfo{pages}{061302} (\bibinfo{year}{1999}), \eprint{astro-ph/9811144}.

\bibitem[{\citenamefont{Fuller et~al.}(2003)\citenamefont{Fuller, Kusenko,
  Mocioiu, and Pascoli}}]{Fuller:2003gy}
\bibinfo{author}{\bibfnamefont{G.~M.} \bibnamefont{Fuller}},
  \bibinfo{author}{\bibfnamefont{A.}~\bibnamefont{Kusenko}},
  \bibinfo{author}{\bibfnamefont{I.}~\bibnamefont{Mocioiu}}, \bibnamefont{and}
  \bibinfo{author}{\bibfnamefont{S.}~\bibnamefont{Pascoli}},
  \bibinfo{journal}{Phys. Rev.} \textbf{\bibinfo{volume}{D68}},
  \bibinfo{pages}{103002} (\bibinfo{year}{2003}), \eprint{astro-ph/0307267}.

\bibitem[{\citenamefont{Kusenko}(2004)}]{Kusenko:2004mm}
\bibinfo{author}{\bibfnamefont{A.}~\bibnamefont{Kusenko}},
  \bibinfo{journal}{Int. J. Mod. Phys.} \textbf{\bibinfo{volume}{D13}},
  \bibinfo{pages}{2065} (\bibinfo{year}{2004}), \eprint{astro-ph/0409521}.

\bibitem[{\citenamefont{Hidaka and Fuller}(2006)}]{Hidaka:2006sg}
\bibinfo{author}{\bibfnamefont{J.}~\bibnamefont{Hidaka}} \bibnamefont{and}
  \bibinfo{author}{\bibfnamefont{G.~M.} \bibnamefont{Fuller}},
  \bibinfo{journal}{Phys. Rev.} \textbf{\bibinfo{volume}{D74}},
  \bibinfo{pages}{125015} (\bibinfo{year}{2006}), \eprint{astro-ph/0609425}.

\bibitem[{\citenamefont{Biermann and Kusenko}(2006)}]{Biermann:2006bu}
\bibinfo{author}{\bibfnamefont{P.~L.} \bibnamefont{Biermann}} \bibnamefont{and}
  \bibinfo{author}{\bibfnamefont{A.}~\bibnamefont{Kusenko}},
  \bibinfo{journal}{Phys. Rev. Lett.} \textbf{\bibinfo{volume}{96}},
  \bibinfo{pages}{091301} (\bibinfo{year}{2006}), \eprint{astro-ph/0601004}.

\bibitem[{\citenamefont{Dodelson and Widrow}(1994)}]{Dodelson:1993je}
\bibinfo{author}{\bibfnamefont{S.}~\bibnamefont{Dodelson}} \bibnamefont{and}
  \bibinfo{author}{\bibfnamefont{L.~M.} \bibnamefont{Widrow}},
  \bibinfo{journal}{Phys. Rev. Lett.} \textbf{\bibinfo{volume}{72}},
  \bibinfo{pages}{17} (\bibinfo{year}{1994}), \eprint{hep-ph/9303287}.

\bibitem[{\citenamefont{Shi and Fuller}(1999)}]{Shi:1998km}
\bibinfo{author}{\bibfnamefont{X.-d.} \bibnamefont{Shi}} \bibnamefont{and}
  \bibinfo{author}{\bibfnamefont{G.~M.} \bibnamefont{Fuller}},
  \bibinfo{journal}{Phys. Rev. Lett.} \textbf{\bibinfo{volume}{82}},
  \bibinfo{pages}{2832} (\bibinfo{year}{1999}), \eprint{astro-ph/9810076}.

\bibitem[{\citenamefont{Abazajian}(2006{\natexlab{a}})}]{AbazajianProduction05}
\bibinfo{author}{\bibfnamefont{K.}~\bibnamefont{Abazajian}},
  \bibinfo{journal}{Phys. Rev.} \textbf{\bibinfo{volume}{D73}},
  \bibinfo{pages}{063506} (\bibinfo{year}{2006}{\natexlab{a}}),
  \eprint{astro-ph/0511630}.

\bibitem[{\citenamefont{Abazajian and Fuller}(2002)}]{Abazajian:2002yz}
\bibinfo{author}{\bibfnamefont{K.~N.} \bibnamefont{Abazajian}}
  \bibnamefont{and} \bibinfo{author}{\bibfnamefont{G.~M.}
  \bibnamefont{Fuller}}, \bibinfo{journal}{Phys. Rev.}
  \textbf{\bibinfo{volume}{D66}}, \bibinfo{pages}{023526}
  (\bibinfo{year}{2002}), \eprint{astro-ph/0204293}.

\bibitem[{\citenamefont{Abazajian
  et~al.}(2001{\natexlab{a}})\citenamefont{Abazajian, Fuller, and
  Patel}}]{Abazajian:2001nj}
\bibinfo{author}{\bibfnamefont{K.}~\bibnamefont{Abazajian}},
  \bibinfo{author}{\bibfnamefont{G.~M.} \bibnamefont{Fuller}},
  \bibnamefont{and} \bibinfo{author}{\bibfnamefont{M.}~\bibnamefont{Patel}},
  \bibinfo{journal}{Phys. Rev.} \textbf{\bibinfo{volume}{D64}},
  \bibinfo{pages}{023501} (\bibinfo{year}{2001}{\natexlab{a}}),
  \eprint{astro-ph/0101524}.

\bibitem[{\citenamefont{Shaposhnikov and Tkachev}(2006)}]{Shaposhnikov:2006xi}
\bibinfo{author}{\bibfnamefont{M.}~\bibnamefont{Shaposhnikov}}
  \bibnamefont{and} \bibinfo{author}{\bibfnamefont{I.}~\bibnamefont{Tkachev}},
  \bibinfo{journal}{Phys. Lett.} \textbf{\bibinfo{volume}{B639}},
  \bibinfo{pages}{414} (\bibinfo{year}{2006}), \eprint{hep-ph/0604236}.

\bibitem[{\citenamefont{Asaka et~al.}(2006)\citenamefont{Asaka, Kusenko, and
  Shaposhnikov}}]{Asaka:2006ek}
\bibinfo{author}{\bibfnamefont{T.}~\bibnamefont{Asaka}},
  \bibinfo{author}{\bibfnamefont{A.}~\bibnamefont{Kusenko}}, \bibnamefont{and}
  \bibinfo{author}{\bibfnamefont{M.}~\bibnamefont{Shaposhnikov}},
  \bibinfo{journal}{Phys. Lett.} \textbf{\bibinfo{volume}{B638}},
  \bibinfo{pages}{401} (\bibinfo{year}{2006}), \eprint{hep-ph/0602150}.

\bibitem[{\citenamefont{Gelmini et~al.}(2004)\citenamefont{Gelmini,
  Palomares-Ruiz, and Pascoli}}]{Gelmini:2004ah}
\bibinfo{author}{\bibfnamefont{G.}~\bibnamefont{Gelmini}},
  \bibinfo{author}{\bibfnamefont{S.}~\bibnamefont{Palomares-Ruiz}},
  \bibnamefont{and} \bibinfo{author}{\bibfnamefont{S.}~\bibnamefont{Pascoli}},
  \bibinfo{journal}{Phys. Rev. Lett.} \textbf{\bibinfo{volume}{93}},
  \bibinfo{pages}{081302} (\bibinfo{year}{2004}), \eprint{astro-ph/0403323}.

\bibitem[{\citenamefont{Abazajian and Koushiappas}(2006)}]{Abazajian:2006yn}
\bibinfo{author}{\bibfnamefont{K.}~\bibnamefont{Abazajian}} \bibnamefont{and}
  \bibinfo{author}{\bibfnamefont{S.~M.} \bibnamefont{Koushiappas}},
  \bibinfo{journal}{Phys. Rev.} \textbf{\bibinfo{volume}{D74}},
  \bibinfo{pages}{023527} (\bibinfo{year}{2006}), \eprint{astro-ph/0605271}.

\bibitem[{\citenamefont{Abazajian
  et~al.}(2001{\natexlab{b}})\citenamefont{Abazajian, Fuller, and
  Tucker}}]{Abazajian:2001vt}
\bibinfo{author}{\bibfnamefont{K.}~\bibnamefont{Abazajian}},
  \bibinfo{author}{\bibfnamefont{G.~M.} \bibnamefont{Fuller}},
  \bibnamefont{and} \bibinfo{author}{\bibfnamefont{W.~H.}
  \bibnamefont{Tucker}}, \bibinfo{journal}{Astrophys. J.}
  \textbf{\bibinfo{volume}{562}}, \bibinfo{pages}{593}
  (\bibinfo{year}{2001}{\natexlab{b}}), \eprint{astro-ph/0106002}.

\bibitem[{\citenamefont{Dolgov and Hansen}(2002)}]{Dolgov:2000ew}
\bibinfo{author}{\bibfnamefont{A.~D.} \bibnamefont{Dolgov}} \bibnamefont{and}
  \bibinfo{author}{\bibfnamefont{S.~H.} \bibnamefont{Hansen}},
  \bibinfo{journal}{Astropart. Phys.} \textbf{\bibinfo{volume}{16}},
  \bibinfo{pages}{339} (\bibinfo{year}{2002}), \eprint{hep-ph/0009083}.

\bibitem[{\citenamefont{Drees and Wright}(2000)}]{Drees:2000wi}
\bibinfo{author}{\bibfnamefont{M.}~\bibnamefont{Drees}} \bibnamefont{and}
  \bibinfo{author}{\bibfnamefont{D.}~\bibnamefont{Wright}}
  (\bibinfo{year}{2000}), \eprint{hep-ph/0006274}.

\bibitem[{\citenamefont{Pal and Wolfenstein}(1982)}]{Pal:1981rm}
\bibinfo{author}{\bibfnamefont{P.~B.} \bibnamefont{Pal}} \bibnamefont{and}
  \bibinfo{author}{\bibfnamefont{L.}~\bibnamefont{Wolfenstein}},
  \bibinfo{journal}{Phys. Rev.} \textbf{\bibinfo{volume}{D25}},
  \bibinfo{pages}{766} (\bibinfo{year}{1982}).

\bibitem[{\citenamefont{Boyarsky
  et~al.}(2006{\natexlab{a}})\citenamefont{Boyarsky, Neronov, Ruchayskiy,
  Shaposhnikov, and Tkachev}}]{Boyarsky:2006fg}
\bibinfo{author}{\bibfnamefont{A.}~\bibnamefont{Boyarsky}},
  \bibinfo{author}{\bibfnamefont{A.}~\bibnamefont{Neronov}},
  \bibinfo{author}{\bibfnamefont{O.}~\bibnamefont{Ruchayskiy}},
  \bibinfo{author}{\bibfnamefont{M.}~\bibnamefont{Shaposhnikov}},
  \bibnamefont{and} \bibinfo{author}{\bibfnamefont{I.}~\bibnamefont{Tkachev}}
  (\bibinfo{year}{2006}{\natexlab{a}}), \eprint{astro-ph/0603660}.

\bibitem[{\citenamefont{Watson et~al.}(2006)\citenamefont{Watson, Beacom,
  Yuksel, and Walker}}]{Watson:2006qb}
\bibinfo{author}{\bibfnamefont{C.~R.} \bibnamefont{Watson}},
  \bibinfo{author}{\bibfnamefont{J.~F.} \bibnamefont{Beacom}},
  \bibinfo{author}{\bibfnamefont{H.}~\bibnamefont{Yuksel}}, \bibnamefont{and}
  \bibinfo{author}{\bibfnamefont{T.~P.} \bibnamefont{Walker}},
  \bibinfo{journal}{Phys. Rev.} \textbf{\bibinfo{volume}{D74}},
  \bibinfo{pages}{033009} (\bibinfo{year}{2006}), \eprint{astro-ph/0605424}.

\bibitem[{\citenamefont{Boyarsky
  et~al.}(2006{\natexlab{b}})\citenamefont{Boyarsky, Nevalainen, and
  Ruchayskiy}}]{Boyarsky:2006ag}
\bibinfo{author}{\bibfnamefont{A.}~\bibnamefont{Boyarsky}},
  \bibinfo{author}{\bibfnamefont{J.}~\bibnamefont{Nevalainen}},
  \bibnamefont{and}
  \bibinfo{author}{\bibfnamefont{O.}~\bibnamefont{Ruchayskiy}}
  (\bibinfo{year}{2006}{\natexlab{b}}), \eprint{astro-ph/0610961}.

\bibitem[{\citenamefont{Riemer-Sorensen
  et~al.}(2006)\citenamefont{Riemer-Sorensen, Hansen, and
  Pedersen}}]{Riemer-Sorensen:2006fh}
\bibinfo{author}{\bibfnamefont{S.}~\bibnamefont{Riemer-Sorensen}},
  \bibinfo{author}{\bibfnamefont{S.~H.} \bibnamefont{Hansen}},
  \bibnamefont{and} \bibinfo{author}{\bibfnamefont{K.}~\bibnamefont{Pedersen}},
  \bibinfo{journal}{Astrophys. J.} \textbf{\bibinfo{volume}{644}},
  \bibinfo{pages}{L33} (\bibinfo{year}{2006}), \eprint{astro-ph/0603661}.

\bibitem[{\citenamefont{McDonald et~al.}(2006)}]{McDonald:2004eu}
\bibinfo{author}{\bibfnamefont{P.}~\bibnamefont{McDonald}}
  \bibnamefont{et~al.}, \bibinfo{journal}{Astrophys. J. Suppl.}
  \textbf{\bibinfo{volume}{163}}, \bibinfo{pages}{80} (\bibinfo{year}{2006}),
  \eprint{astro-ph/0405013}.

\bibitem[{\citenamefont{Seljak et~al.}(2006{\natexlab{a}})\citenamefont{Seljak,
  Makarov, McDonald, and Trac}}]{Seljak:2006qw}
\bibinfo{author}{\bibfnamefont{U.}~\bibnamefont{Seljak}},
  \bibinfo{author}{\bibfnamefont{A.}~\bibnamefont{Makarov}},
  \bibinfo{author}{\bibfnamefont{P.}~\bibnamefont{McDonald}}, \bibnamefont{and}
  \bibinfo{author}{\bibfnamefont{H.}~\bibnamefont{Trac}},
  \bibinfo{journal}{Phys. Rev. Lett.} \textbf{\bibinfo{volume}{97}},
  \bibinfo{pages}{191303} (\bibinfo{year}{2006}{\natexlab{a}}),
  \eprint{astro-ph/0602430}.

\bibitem[{\citenamefont{Viel et~al.}(2006)\citenamefont{Viel, Lesgourgues,
  Haehnelt, Matarrese, and Riotto}}]{Viel:2006kd}
\bibinfo{author}{\bibfnamefont{M.}~\bibnamefont{Viel}},
  \bibinfo{author}{\bibfnamefont{J.}~\bibnamefont{Lesgourgues}},
  \bibinfo{author}{\bibfnamefont{M.~G.} \bibnamefont{Haehnelt}},
  \bibinfo{author}{\bibfnamefont{S.}~\bibnamefont{Matarrese}},
  \bibnamefont{and} \bibinfo{author}{\bibfnamefont{A.}~\bibnamefont{Riotto}},
  \bibinfo{journal}{Phys. Rev. Lett.} \textbf{\bibinfo{volume}{97}},
  \bibinfo{pages}{071301} (\bibinfo{year}{2006}), \eprint{astro-ph/0605706}.

\bibitem[{\citenamefont{Viel et~al.}(2005)\citenamefont{Viel, Lesgourgues,
  Haehnelt, Matarrese, and Riotto}}]{Viel:2005qj}
\bibinfo{author}{\bibfnamefont{M.}~\bibnamefont{Viel}},
  \bibinfo{author}{\bibfnamefont{J.}~\bibnamefont{Lesgourgues}},
  \bibinfo{author}{\bibfnamefont{M.~G.} \bibnamefont{Haehnelt}},
  \bibinfo{author}{\bibfnamefont{S.}~\bibnamefont{Matarrese}},
  \bibnamefont{and} \bibinfo{author}{\bibfnamefont{A.}~\bibnamefont{Riotto}},
  \bibinfo{journal}{Phys. Rev.} \textbf{\bibinfo{volume}{D71}},
  \bibinfo{pages}{063534} (\bibinfo{year}{2005}), \eprint{astro-ph/0501562}.

\bibitem[{\citenamefont{Abazajian}(2006{\natexlab{b}})}]{Abazajian:2005xn}
\bibinfo{author}{\bibfnamefont{K.}~\bibnamefont{Abazajian}},
  \bibinfo{journal}{Phys. Rev.} \textbf{\bibinfo{volume}{D73}},
  \bibinfo{pages}{063513} (\bibinfo{year}{2006}{\natexlab{b}}),
  \eprint{astro-ph/0512631}.

\bibitem[{\citenamefont{Seljak et~al.}(2006{\natexlab{b}})\citenamefont{Seljak,
  Slosar, and McDonald}}]{Seljak:2006bg}
\bibinfo{author}{\bibfnamefont{U.}~\bibnamefont{Seljak}},
  \bibinfo{author}{\bibfnamefont{A.}~\bibnamefont{Slosar}}, \bibnamefont{and}
  \bibinfo{author}{\bibfnamefont{P.}~\bibnamefont{McDonald}},
  \bibinfo{journal}{JCAP} \textbf{\bibinfo{volume}{0610}}, \bibinfo{pages}{014}
  (\bibinfo{year}{2006}{\natexlab{b}}), \eprint{astro-ph/0604335}.

\bibitem[{\citenamefont{Mangano et~al.}(2005)}]{Mangano:2005cc}
\bibinfo{author}{\bibfnamefont{G.}~\bibnamefont{Mangano}} \bibnamefont{et~al.},
  \bibinfo{journal}{Nucl. Phys.} \textbf{\bibinfo{volume}{B729}},
  \bibinfo{pages}{221} (\bibinfo{year}{2005}), \eprint{hep-ph/0506164}.

\bibitem[{\citenamefont{Cyburt et~al.}(2005)\citenamefont{Cyburt, Fields,
  Olive, and Skillman}}]{Cyburt:2004yc}
\bibinfo{author}{\bibfnamefont{R.~H.} \bibnamefont{Cyburt}},
  \bibinfo{author}{\bibfnamefont{B.~D.} \bibnamefont{Fields}},
  \bibinfo{author}{\bibfnamefont{K.~A.} \bibnamefont{Olive}}, \bibnamefont{and}
  \bibinfo{author}{\bibfnamefont{E.}~\bibnamefont{Skillman}},
  \bibinfo{journal}{Astropart. Phys.} \textbf{\bibinfo{volume}{23}},
  \bibinfo{pages}{313} (\bibinfo{year}{2005}), \eprint{astro-ph/0408033}.

\bibitem[{\citenamefont{Hui et~al.}(2006)}]{Hui}
\bibinfo{author}{\bibfnamefont{L.}~\bibnamefont{Hui}} \bibnamefont{et~al.}
  (\bibinfo{year}{2006}), \eprint{in preparation}.

\bibitem[{\citenamefont{Hickox and Markevitch}(2006)}]{Hickox:2005dz}
\bibinfo{author}{\bibfnamefont{R.~C.} \bibnamefont{Hickox}} \bibnamefont{and}
  \bibinfo{author}{\bibfnamefont{M.}~\bibnamefont{Markevitch}},
  \bibinfo{journal}{Astrophys. J.} \textbf{\bibinfo{volume}{645}},
  \bibinfo{pages}{95} (\bibinfo{year}{2006}), \eprint{astro-ph/0512542}.

\bibitem[{\citenamefont{Hickox and Markevitch}(2007)}]{HickoxInPrep}
\bibinfo{author}{\bibfnamefont{R.~C.} \bibnamefont{Hickox}} \bibnamefont{and}
  \bibinfo{author}{\bibfnamefont{M.}~\bibnamefont{Markevitch}},
  \bibinfo{journal}{in prep.}  (\bibinfo{year}{2007}).

\bibitem[{\citenamefont{McCammon et~al.}(2002)}]{McCammon:2002gb}
\bibinfo{author}{\bibfnamefont{D.}~\bibnamefont{McCammon}}
  \bibnamefont{et~al.}, \bibinfo{journal}{Astrophys. J.}
  \textbf{\bibinfo{volume}{576}}, \bibinfo{pages}{188} (\bibinfo{year}{2002}),
  \eprint{astro-ph/0205012}.

\bibitem[{\citenamefont{Barger et~al.}(1995)\citenamefont{Barger, Phillips, and
  Sarkar}}]{Barger:1995ty}
\bibinfo{author}{\bibfnamefont{V.~D.} \bibnamefont{Barger}},
  \bibinfo{author}{\bibfnamefont{R.~J.~N.} \bibnamefont{Phillips}},
  \bibnamefont{and} \bibinfo{author}{\bibfnamefont{S.}~\bibnamefont{Sarkar}},
  \bibinfo{journal}{Phys. Lett.} \textbf{\bibinfo{volume}{B352}},
  \bibinfo{pages}{365} (\bibinfo{year}{1995}), \eprint{hep-ph/9503295}.

\bibitem[{\citenamefont{Asaka et~al.}(2007)\citenamefont{Asaka, Laine, and
  Shaposhnikov}}]{Asaka:2006nq}
\bibinfo{author}{\bibfnamefont{T.}~\bibnamefont{Asaka}},
  \bibinfo{author}{\bibfnamefont{M.}~\bibnamefont{Laine}}, \bibnamefont{and}
  \bibinfo{author}{\bibfnamefont{M.}~\bibnamefont{Shaposhnikov}},
  \bibinfo{journal}{JHEP} \textbf{\bibinfo{volume}{01}}, \bibinfo{pages}{091}
  (\bibinfo{year}{2007}), \eprint{hep-ph/0612182}.

\bibitem[{\citenamefont{{Klypin} et~al.}(2002)\citenamefont{{Klypin}, {Zhao},
  and {Somerville}}}]{Klypin2001}
\bibinfo{author}{\bibfnamefont{A.}~\bibnamefont{{Klypin}}},
  \bibinfo{author}{\bibfnamefont{H.}~\bibnamefont{{Zhao}}}, \bibnamefont{and}
  \bibinfo{author}{\bibfnamefont{R.~S.} \bibnamefont{{Somerville}}},
  \bibinfo{journal}{\apj} \textbf{\bibinfo{volume}{573}}, \bibinfo{pages}{597}
  (\bibinfo{year}{2002}), \eprint{astro-ph/0110390}.

\bibitem[{\citenamefont{Battaglia et~al.}(2005)}]{Battaglia:2005rj}
\bibinfo{author}{\bibfnamefont{G.}~\bibnamefont{Battaglia}}
  \bibnamefont{et~al.}, \bibinfo{journal}{Mon. Not. Roy. Astron. Soc.}
  \textbf{\bibinfo{volume}{364}}, \bibinfo{pages}{433} (\bibinfo{year}{2005}),
  \eprint{astro-ph/0506102}.

\bibitem[{\citenamefont{Widrow and Dubinski}(2005)}]{Widrow:2005bt}
\bibinfo{author}{\bibfnamefont{L.~M.} \bibnamefont{Widrow}} \bibnamefont{and}
  \bibinfo{author}{\bibfnamefont{J.}~\bibnamefont{Dubinski}},
  \bibinfo{journal}{Astrophys. J.} \textbf{\bibinfo{volume}{631}},
  \bibinfo{pages}{838} (\bibinfo{year}{2005}), \eprint{astro-ph/0506177}.

\bibitem[{\citenamefont{Navarro et~al.}(1996)\citenamefont{Navarro, Frenk, and
  White}}]{NFW96}
\bibinfo{author}{\bibfnamefont{J.~F.} \bibnamefont{Navarro}},
  \bibinfo{author}{\bibfnamefont{C.~S.} \bibnamefont{Frenk}}, \bibnamefont{and}
  \bibinfo{author}{\bibfnamefont{S.~D.~M.} \bibnamefont{White}},
  \bibinfo{journal}{Astrophys. J.} \textbf{\bibinfo{volume}{462}},
  \bibinfo{pages}{563} (\bibinfo{year}{1996}), \eprint{astro-ph/9508025}.

\bibitem[{\citenamefont{{Avila-Reese} et~al.}(2001)\citenamefont{{Avila-Reese},
  {Col{\'{\i}}n}, {Valenzuela}, {D'Onghia}, and
  {Firmani}}}]{Avila-Reese:2000hg}
\bibinfo{author}{\bibfnamefont{V.}~\bibnamefont{{Avila-Reese}}},
  \bibinfo{author}{\bibfnamefont{P.}~\bibnamefont{{Col{\'{\i}}n}}},
  \bibinfo{author}{\bibfnamefont{O.}~\bibnamefont{{Valenzuela}}},
  \bibinfo{author}{\bibfnamefont{E.}~\bibnamefont{{D'Onghia}}},
  \bibnamefont{and}
  \bibinfo{author}{\bibfnamefont{C.}~\bibnamefont{{Firmani}}},
  \bibinfo{journal}{\apj} \textbf{\bibinfo{volume}{559}}, \bibinfo{pages}{516}
  (\bibinfo{year}{2001}), \eprint{astro-ph/0010525}.

\bibitem[{\citenamefont{Weisskopf et~al.}(2002)}]{Weisskopf:2001uu}
\bibinfo{author}{\bibfnamefont{M.~C.} \bibnamefont{Weisskopf}}
  \bibnamefont{et~al.}, \bibinfo{journal}{Publ. Astron. Soc. Pac.}
  \textbf{\bibinfo{volume}{114}}, \bibinfo{pages}{1} (\bibinfo{year}{2002}),
  \eprint{astro-ph/0110308}.

\bibitem[{\citenamefont{Brandt et~al.}(2001)}]{Brandt:2001vb}
\bibinfo{author}{\bibfnamefont{W.~N.} \bibnamefont{Brandt}}
  \bibnamefont{et~al.}, \bibinfo{journal}{Astrophys. J.}
  \textbf{\bibinfo{volume}{122}}, \bibinfo{pages}{1} (\bibinfo{year}{2001}),
  \eprint{astro-ph/0102411}.

\bibitem[{\citenamefont{Giacconi et~al.}(2002)}]{Giacconi:2001tb}
\bibinfo{author}{\bibfnamefont{R.}~\bibnamefont{Giacconi}}
  \bibnamefont{et~al.}, \bibinfo{journal}{Astrophys. J. Suppl.}
  \textbf{\bibinfo{volume}{139}}, \bibinfo{pages}{369} (\bibinfo{year}{2002}),
  \eprint{astro-ph/0112184}.

\bibitem[{\citenamefont{Smith et~al.}(2001)\citenamefont{Smith, Brickhouse,
  Liedahl, and Raymond}}]{Smith:2001he}
\bibinfo{author}{\bibfnamefont{R.~K.} \bibnamefont{Smith}},
  \bibinfo{author}{\bibfnamefont{N.~S.} \bibnamefont{Brickhouse}},
  \bibinfo{author}{\bibfnamefont{D.~A.} \bibnamefont{Liedahl}},
  \bibnamefont{and} \bibinfo{author}{\bibfnamefont{J.~C.}
  \bibnamefont{Raymond}}, \bibinfo{journal}{Astrophys. J.}
  \textbf{\bibinfo{volume}{556}}, \bibinfo{pages}{L91} (\bibinfo{year}{2001}),
  \eprint{astro-ph/0106478}.

\bibitem[{\citenamefont{{Protassov} et~al.}(2002)\citenamefont{{Protassov},
  {van Dyk}, {Connors}, {Kashyap}, and {Siemiginowska}}}]{Protassov:2002sz}
\bibinfo{author}{\bibfnamefont{R.}~\bibnamefont{{Protassov}}},
  \bibinfo{author}{\bibfnamefont{D.~A.} \bibnamefont{{van Dyk}}},
  \bibinfo{author}{\bibfnamefont{A.}~\bibnamefont{{Connors}}},
  \bibinfo{author}{\bibfnamefont{V.~L.} \bibnamefont{{Kashyap}}},
  \bibnamefont{and}
  \bibinfo{author}{\bibfnamefont{A.}~\bibnamefont{{Siemiginowska}}},
  \bibinfo{journal}{\apj} \textbf{\bibinfo{volume}{571}}, \bibinfo{pages}{545}
  (\bibinfo{year}{2002}), \eprint{astro-ph/0201547}.

\bibitem[{\citenamefont{Boyarsky
  et~al.}(2006{\natexlab{c}})\citenamefont{Boyarsky, Neronov, Ruchayskiy, and
  Shaposhnikov}}]{Boyarsky:2005us}
\bibinfo{author}{\bibfnamefont{A.}~\bibnamefont{Boyarsky}},
  \bibinfo{author}{\bibfnamefont{A.}~\bibnamefont{Neronov}},
  \bibinfo{author}{\bibfnamefont{O.}~\bibnamefont{Ruchayskiy}},
  \bibnamefont{and}
  \bibinfo{author}{\bibfnamefont{M.}~\bibnamefont{Shaposhnikov}},
  \bibinfo{journal}{Mon. Not. Roy. Astron. Soc.}
  \textbf{\bibinfo{volume}{370}}, \bibinfo{pages}{213}
  (\bibinfo{year}{2006}{\natexlab{c}}), \eprint{astro-ph/0512509}.

\bibitem[{\citenamefont{Boyarsky
  et~al.}(2006{\natexlab{d}})\citenamefont{Boyarsky, Neronov, Ruchayskiy, and
  Shaposhnikov}}]{Boyarsky:2006zi}
\bibinfo{author}{\bibfnamefont{A.}~\bibnamefont{Boyarsky}},
  \bibinfo{author}{\bibfnamefont{A.}~\bibnamefont{Neronov}},
  \bibinfo{author}{\bibfnamefont{O.}~\bibnamefont{Ruchayskiy}},
  \bibnamefont{and}
  \bibinfo{author}{\bibfnamefont{M.}~\bibnamefont{Shaposhnikov}},
  \bibinfo{journal}{Phys. Rev.} \textbf{\bibinfo{volume}{D74}},
  \bibinfo{pages}{103506} (\bibinfo{year}{2006}{\natexlab{d}}),
  \eprint{astro-ph/0603368}.

\end{thebibliography}

\end{document}